\documentclass[aps,prl,twocolumn,showpacs,superscriptaddress,groupedaddress,nobalancelastpage]{revtex4}  
\usepackage{graphicx}  
\usepackage{dcolumn}   
\usepackage{bm}        
\usepackage{amssymb}   
\usepackage{epstopdf}

\begin{document}

\widetext


\title{Nonlinear Resonances and Antiresonances of a Forced Sonic Vacuum}
\author{D. Pozharskiy}
\affiliation{Department of Chemical and Biological Engineering and PACM, Princeton University, Princeton, NJ 08544, USA}

\author{Y. Zhang}
\affiliation{Department of Mechanical Science and Engineering, University
of Illinois at Urbana-Champaign, 1206 West Green Street, Urbana, IL
61822, USA}

\author{M.O. Williams}
\affiliation{Department of Chemical and Biological Engineering and PACM, Princeton University, Princeton, NJ 08544, USA}

\author{D.M. McFarland}
\affiliation{Department of Aerospace Engineering, University of Illinois at Urbana - Champaign, Urbana, IL 61822, USA}

\author{P.G. Kevrekidis}
\affiliation{Department of Mathematics and Statistics, University of Massachusetts,
Amherst, MA 01003-4515, USA}

\affiliation{Center for Nonlinear Studies and Theoretical Division, Los Alamos
National Laboratory, Los Alamos, NM 87544, USA}

\author{A.F. Vakakis}
\affiliation{Department of Mechanical Science and Engineering, University
of Illinois at Urbana-Champaign, 1206 West Green Street, Urbana, IL
61822, USA}

\author{I.G. Kevrekidis}
\affiliation{Department of Chemical and Biological Engineering
and PACM, Princeton University, Princeton, NJ 08544, USA}

\date{\today}

\begin{abstract}
We consider a harmonically driven acoustic medium in the
form of a (finite length) highly nonlinear
granular crystal with an amplitude and frequency dependent
boundary drive.
Remarkably, despite the absence of a linear spectrum
in the system, we identify resonant periodic propagation whereby
the crystal responds at {\it integer multiples} of the drive period, and
observe that they can
lead to local maxima of transmitted force at its fixed boundary.
In addition, we identify and discuss minima of the transmitted force
(``antiresonances") in between these resonances.
Representative one-parameter
complex bifurcation diagrams involve period doublings,
Neimark-Sacker bifurcations as well
as multiple isolas (e.g. of period-3, -4 or -5 solutions entrained by the forcing).
We combine them in a more detailed,  two-parameter bifurcation
diagram describing the stability
of such responses  to both frequency and amplitude variations of the drive.
This picture supports an unprecedented example of a notion of
a (purely) ``nonlinear spectrum'' in a system which allows no sound wave propagation
(due to zero sound speed: the so-called sonic vacuum).
We rationalize this behavior in terms
of purely nonlinear building blocks: apparent traveling and standing
nonlinear waves.
\end{abstract}

\maketitle

{\it Introduction.} In media that bear a crystalline lattice structure
(perfect or imperfect), an understanding of the underlying linear
spectrum is of paramount importance both for their linear wave
transmission properties~\cite{maradudin} and for the emergence
of anharmonicity-induced excitations such as discrete breathers~\cite{flach}.
However, in recent years there has been an increased interest in
systems that {\em do not possess a linear spectrum}.
This
stems to a considerable degree from the interest in discretizing
prototypical nonlinear systems such as the Burgers equation~\cite{bennaim},
but also from the intrinsic mathematical (and potentially physical)
interest in exploring systems with purely nonlinear dispersion;
a remarkable feature of the latter  has been the
identification of discrete analogues of
compactons~\cite{rosenau} and of compactly
supported breathers~\cite{flach2}.
A fundamental question that then emerges
in the latter context is whether an analogue of acoustic transmission/spectral
features exists in these highly (or purely) nonlinear systems.

Importantly, the last two decades have also brought forth a
remarkably accessible (theoretically, numerically and experimentally)
example of a (tunably) highly nonlinear system.
This is in the form of
granular crystals, consisting of lattices of beads
interacting (chiefly) elastically, by means of the so-called
Hertzian contact
forces~\cite{Nesterenko2001,Sen2008} $F= A \delta^p$.
Here $A$ is an elasticity and geometry dependent prefactor,
$\delta$ is the compression-induced displacement and $p$ the nonlinear
force exponent, typically equal to $3/2$.
These acoustic systems enjoy a remarkable degree of tunability
from homogeneous (same types of beads) to heterogeneous or disordered,
and importantly for our current purposes from essentially linear to
weakly or even highly or purely
nonlinear~\cite{Nesterenko2001,Sen2008,Kevrekidis2011,Theocharis_rev}.
The latter aspect is controlled by the amount of {\em a priori} compression
(dubbed precompression $\delta_0$) externally imposed on the beads.
When this
precompression is absent, i.e., $\delta_0=0$, the Hertzian interaction
of the particle
chain manifests its {\it essentially} nonlinear character.
It is
in that case that the speed of sound $c=(A p \delta_0^{p-1})^{1/2}$ in
the medium vanishes, the so-called sonic vacuum scenario.
In this
setting, highly nonlinear solitary waves, pioneered by
Nesterenko~\cite{nesterenko1983,Nesterenko2001}, have been identified
and are the dominant energy carriers in the system~\cite{hinch}.

Our principal question is thus formulated at the junction of these
two important themes. 
On the one hand, for linear systems, acoustic
wave propagation and the linear ``phonon'' spectrum are crucial for
a fundamental understanding of propagating vs. evanescent waves
in the system.
On the other hand, no such entities arise in the
sonic vacuum. 
In the latter case, highly localized (in fact, doubly exponentially
localized~\cite{ahnert}) solitary waves constitute the prototypical
excitation of the system.
Can then one seek
spectral properties/spectral response in such highly nonlinear
(finite lattice) settings and, if the answer is positive, how does such a response emerge?
Surprisingly, in our system of choice (towards exploring these
ideas), namely the granular crystal, we computationally illustrate the existence of such a ``nonlinear spectrum''
as it arises in response to externally driving the system  at different frequencies.
The principal mechanism for the emergence of this spectrum is the (nonlinearly induced) periodic and potentially
quasi-periodic response of the
system depending on the relative phases between the applied harmonic excitation
and waves (``pulses") that appear to propagate in opposite directions, or to form standing wave patterns within the medium.
These inherently nonlinear notions have the potential to be widely applicable
to systems without a linear spectrum,
well beyond the specifics of our particular paradigm, and hence are
of broad and diverse interest.

Only a few works have discussed the responses of granular crystals to harmonic excitation.
In \cite{Lyndon2015} nonlinear resonances in precompressed granular crystals were studied,
together with certain bifurcations related with these motions;
however, in distinction from the system considered in the present work, the systems in~\cite{Lyndon2015} possess
a linear spectrum due to applied static precompression.
In \cite{Lyndon2013} the forced response of a two-bead granular crystal was considered and the pass and stop bands of
this system were confirmed; given, however, the light damping of that system, no clear resonance peaks could be detected since the high-amplitude response became chaotic.
To date the resonance and antiresonance structure of nonlinear sonic vacua have not been systematically studied; this is our aim here.

{\it Setup.} We consider a chain of beads interacting elastically
via Hertzian contacts that are mathematically described (upon a
suitable rescaling of amplitudes and of time~\cite{starosv}) as:
\begin{eqnarray}
\ddot{x}_i &=& (x_{i-1}-x_i)_+^{3/2} - (x_i-x_{i+1})_+^{3/2}
\nonumber
\\
\hspace{2em}&&+ \lambda \left[(\dot{x}_{i-1}-\dot{x}_i) H(x_{i-1}-x_i)\right. \nonumber\\
\hspace{2em}&&\hspace{1em}- \left.(\dot{x}_i-\dot{x}_{i+1}) H(x_i-x_{i+1})\right],
\label{eqn1}
\end{eqnarray}
with $i=1,\dots,N$ with $N=11$).
The $x_i$'s here
represent the displacement of the particle centers from their
equilibrium positions.
The first two terms on the right hand side capture the bead-bead
elastic interaction in our finite lattice, while the last two
account for the linear dissipative dynamics, following the proposal
of~\cite{nester3}.
$H$ denotes the Heaviside function, while $(\cdot)_+ = \max(\cdot, 0)$.
These terms appear because of the tensionless character of the system,
which implies there are no forces when beads are not in contact.
The rescaling used to bring the system in the form (\ref{eqn1}) together with the numerical values of the system parameters are given in the
Appendix.

The boundaries of the system under consideration consist
of (a) an actuator exciting the (zero-th) bead with prescribed harmonic displacement at the left end
and (b) an immobile wall on the right end.
The latter is implemented as a fixed additional bead for which $x_{N+1}=\dot{x}_{N+1}=0$, while
the former is associated with $x_0=A \sin(\omega t)$, i.e., a
periodic excitation of a given frequency $\omega$ with an amplitude $A$.
The result of this
harmonic drive in the presence of {\it linear} coupling between adjacent
nodes of a (finite) lattice would be to excite the quantized (based
on the boundary conditions) wavenumbers allowable by the lattice,
and, accordingly, the frequencies corresponding to these wavenumbers
based on the linear dispersion relation.
In the present case, however, such a linear dispersion relation is {\em completely absent} (and so is a linear spectrum of frequencies) 
due to the essentially nonlinear character
of the examined sonic vacuum.
Accordingly, the resulting dynamics are expected to be
completely tunable with the amplitude of the forcing, whereas the existence and structure of resonances in this system
(i.e., of amplitude amplification on the granular chain at certain frequencies) is an interesting question.
In the absence of linear (or linearized) dynamics, we now seek the corresponding response diagram in our
highly nonlinear chain.

\begin{figure*}
\includegraphics[width=\textwidth]{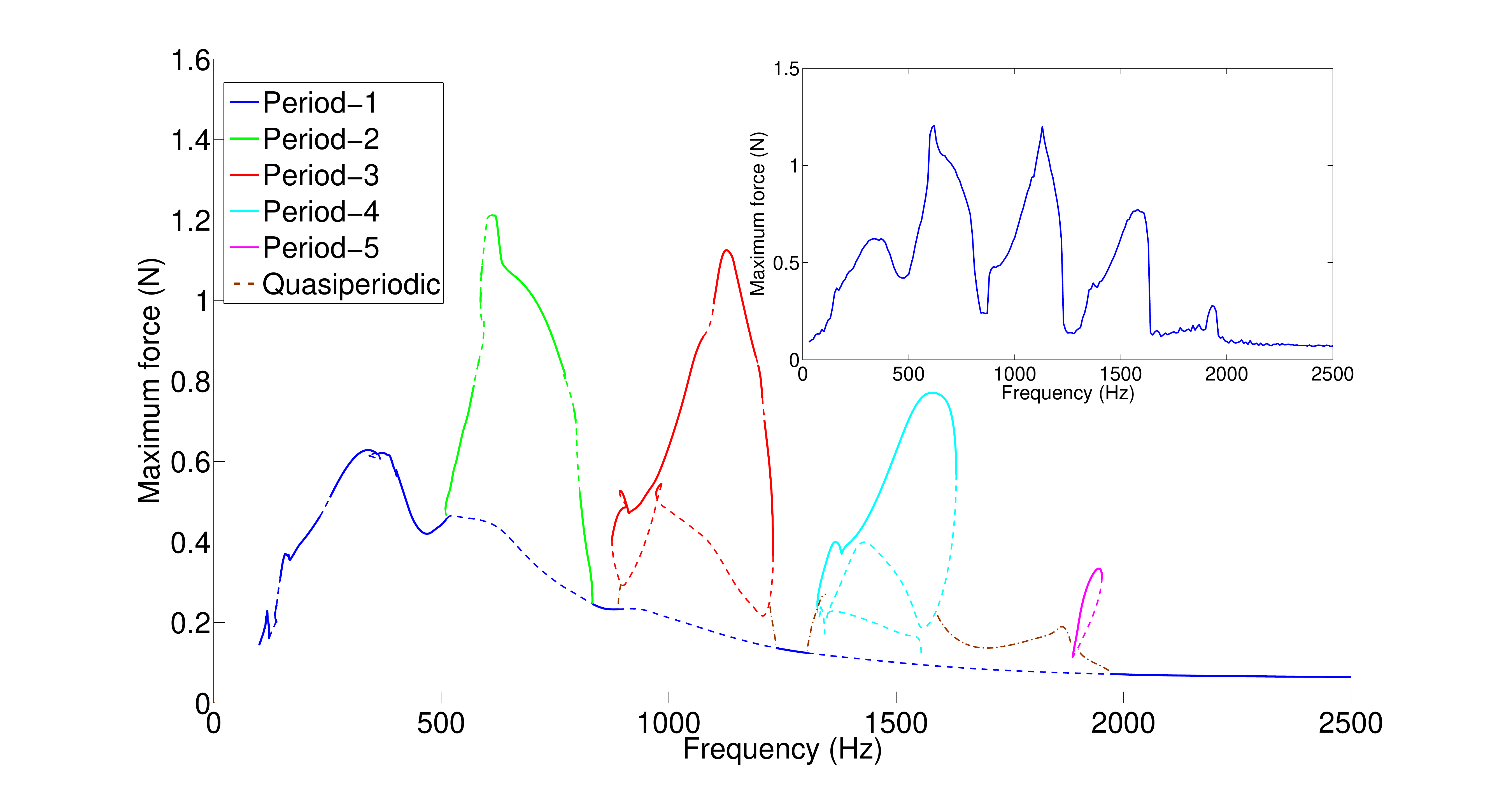}
\caption{(Color Online) Maximum force exerted on the fixed wall by the last bead
over one period of the periodic motion of the crystal. Branches
corresponding to periods that are different integer multiples of the period of the actuator 
are shown in different colors.
In the quasiperiodic case, the motion
becomes dense on a $T^2$ torus, and the average force is estimated over a
long time frame. 
The main graph showcases the different
branches of nonlinear solutions (denoted by solid when stable,
and dashed when unstable), while the inset incorporates solely
the stable, observable  response, i.e., the ``nonlinear spectrum'', that would
be
experimentally measured. 
The amplitude of the periodic excitation used here
is $5\times10^{-7}$ m.}
\label{kvfig1}
\end{figure*}

{\it Results.} The central result is summarized in Fig.~\ref{kvfig1}.
This represents the ``nonlinear spectrum'' of the system in
response to its periodic actuation from the left boundary.
To obtain this diagram, a combination of a fixed point algorithm (Newton's method
for boundary value problems in time) and
pseudoarclength continuation was used to numerically compute the branches of period-$k$ solutions
for various integers $k$.
Similar to \cite{Chong2014}, this was accomplished by identifying fixed points of the stroboscopic map with 
a period of $k 2\pi/\omega$ (i.e., $k$ times the period of the applied harmonic excitation).
%
The inset captures solely the {\em stable} (observable) response
of the system, in the form of the maximal force exerted on the
wall by the last bead over the period of the solution, plotted as a function
of the drive frequency.
It bears five
clearly distinguishable peaks (which we will call 
``resonances'' in the sense of local maxima of the response
of the granular crystal), indicating that at these frequencies
a maximal force is transmitted to the right end of the crystal.
The diagram also features
dips (which we will call ``antiresonances") between the peaks, associated with frequencies that
locally minimize the transmitted force.
This resonance structure
is clearly reminiscent of a linear spectrum, despite the fact that
the basic ingredients of the latter
are absent.
Moreover, the frequency range where these resonances and antiresonances occur lies within the nonlinear pass band of the chain~\cite{Jayaprakash},
i.e., the frequency range where solitary pulses or spatially extended waves can propagate through the corresponding granular crystal of infinite extent,
whereas for frequencies above 2 kHz, there is negligible force transmission.
As we will see below, the {\em effective phase variation} (defined below and in the figure captions) between
the applied excitation and the nonlinear waves that result from it 
and travel along the chain is
crucial in the formation of these persistent resonance dynamics.
We then delve into a more detailed exploration
of these features.

We seek exact periodic orbits for the different frequencies within
the range of our actuator (here, between $0$ and $2.5$ kHz).
The resulting limit cycle solutions (given that this is a driven, damped
system) are shown in the bifurcation diagram
of the main part of Fig.~\ref{kvfig1}.
There, it can be discerned
that there is a ``backbone'' of harmonic response (consisting of period-1
nonlinear solutions, with oscillation frequencies matching the driving frequency).
In addition to identifying the existence of such a branch,
we also explore its stability by means of Floquet analysis,
identifying the corresponding Floquet multipliers of each point; see,
e.g.,~\cite{flach,Chong2014}
for a detailed discussion.
When these multipliers are within the
unit circle, they correspond to stable dynamics, while their
exiting the unit circle is associated with instability, which,
in turn, is denoted by dashed lines in Fig.~\ref{kvfig1}.

While the period-1 solution is stable at low and high frequencies,
indicating a robust dynamical response of the system in these
ranges, it undergoes a series of destabilizing bifurcations in
intermediate frequency regimes, which lead to the observation of stable
branches of higher period (subharmonic) solutions, i.e., of
periodic responses with periods that
are integer multiples of the period of the applied excitation.
These subharmonics are denoted by different colors in Fig.~\ref{kvfig1}.
We note the existence of a resonant peak ``within"
each branch of periodic solutions, corresponding to a
local maximum of the magnitude of the largest (over a single
period of the solution)
transmitted force
to the right end of the chain.
On the contrary, antiresonances corresponding to local minima of
the largest transmitted force typically occur
on stable parts of the period-1 solution branch.
The detailed spatio-temporal nature of the periodic responses of the chain in resonances 
and antiresonances is of
particular interest and will be discussed below.
From Fig.~\ref{kvfig1} we deduce that the emergence of the period-2
solution occurs through a period-doubling bifurcation (and its
disappearance through the ``inverse" of the same bifurcation).
However, the rest of the branches are isolated (so-called ``isolas'')
and are not emerging directly
from the primary 1:1 branch.
It is also observed, however, that there are regions
where we did not find low-period stable periodic solutions; in these regions
(see the brown
dash-dotted lines in Fig.~\ref{kvfig1}), we expect the system response to be
quasi-periodic (i.e., consisting of two harmonics with irrationally related
frequencies) - or periodic with much higher
integer periods, as these are known to arise in a systematic, interspersed
pattern in quasiperiodic regimes~\cite{aronson1982,kevr1986,aronson1986}.

\begin{figure*}
\includegraphics[width=\textwidth]{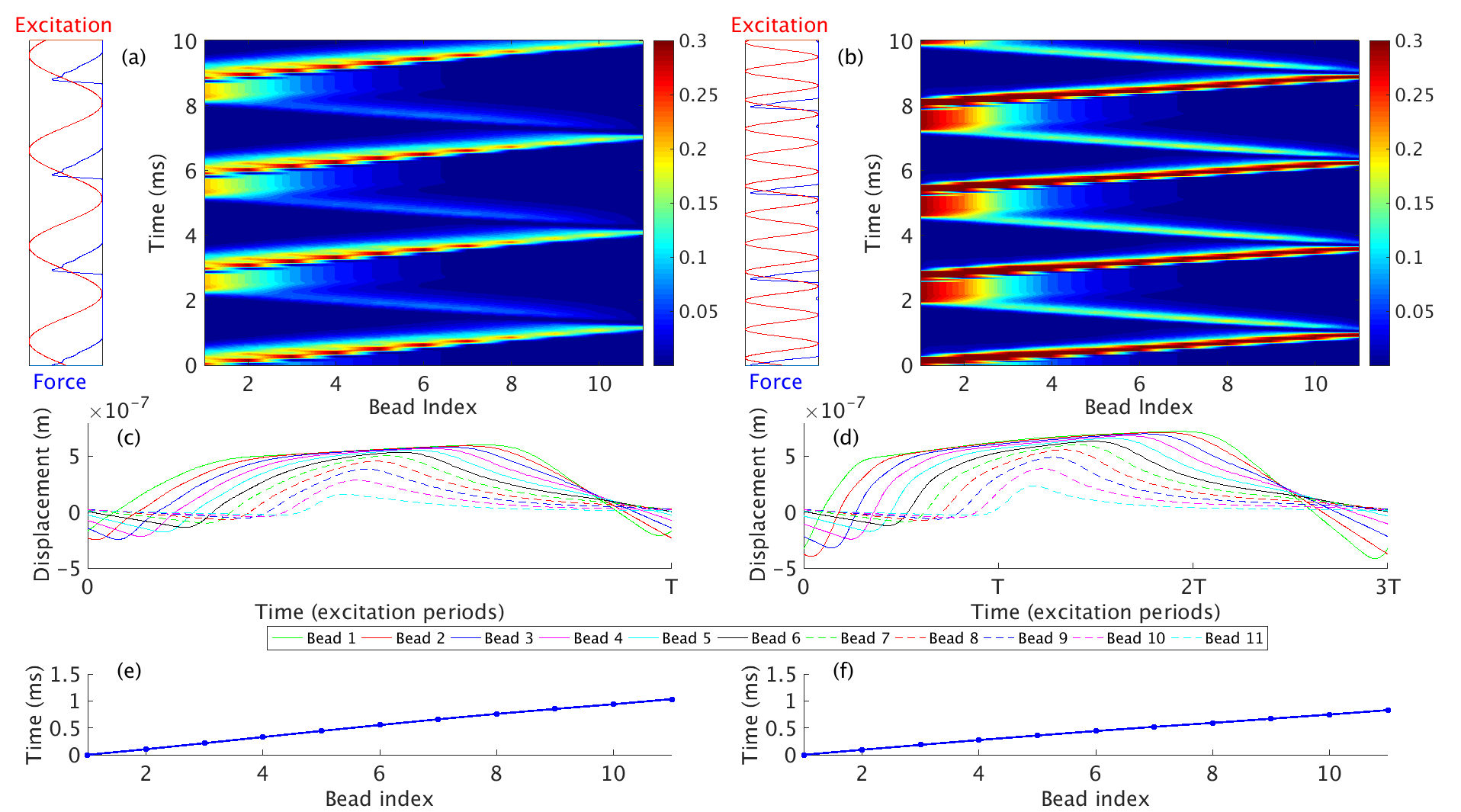}
\caption{(Color Online) (a),(b) Long-term, stationary spatio-temporal variation of the kinetic energy (scaled$\times {10^{7}}$) for (a) 1:1 resonance (340 Hz) and (b) 1:3 resonance (1130 Hz);
the input harmonic displacement and applied force excitations for each case are also shown on the left of each plot. (c),(d) Displacement waveforms of
every bead during one full period of the crystal motion. (e),(f) Effective phase variation of the individual bead responses (computed by the time instants when each bead's velocity
first becomes positive and when the leftmost bead's velocity first becomes positive); this highlights the  
``traveling pulse" nature of the long-term, stationary dynamics of the granular chain at the resonance peaks.}
\label{kvfig2}
\end{figure*}

Figure~\ref{kvfig2} depicts representative spatio-temporal variations of the kinetic energy in the
regime of the 1:1 and 1:3 resonance peaks, together with the corresponding displacement waveforms for every
bead during one period.
Visual inspection of the kinetic energy space-time plots clearly suggests pulses traveling along the chain
(part of the ``nonlinear spectrum building blocks" mentioned in the abstract). 
These apparent waves are initiated on the left
end of the chain, travel to the right, become (partially) reflected there, come back to the left end of the chain,
and there interact again with the next ``hit" of the actuator.
The third component of each figure summarizes this pulse traveling in the form of an effective phase variation in the response of each bead:
what is plotted is the time difference between the first occurrence during a period (at stationarity), of positive velocity for each bead, and
for the leftmost bead of the chain; the almost perfect straight line plot clearly implies the traveling pulse nature of the long-term, stationary dynamics in the resonance peaks considered.
This provides some insight into the nature of the relevant periodic, nonlinear solutions that dominate the response at resonance
of the crystal lattice under the periodic external excitation.

In the parameter regime of the resonance peaks, it appears that
the effective phase variation between the applied excitation and the traveling pulse initiated in the crystal is such that strong energy
transfer from the exciting source occurs upon impact.
This can be rationalized when we take into account that all resonances occur within the nonlinear pass band of the crystal, 
where traveling pulses constitute the
nonlinear mechanism of energy transmission, and the applied
excitation is in the form of a periodic train of impulses~\cite{Nesterenko2001,Sen2008,Kevrekidis2011,Theocharis_rev,Hasan}.
In the case of 1:1 resonance (Fig.~\ref{kvfig2}(a)) a strong impulse excitation
occurs at precisely the time instant when the traveling pulse reflected from the right immovable boundary reaches the left end; 
one can deduce that the traveling pulses fully synchronize with the excitation source.
At higher-order resonances, however,
the time needed by the traveling pulse to fully traverse twice the length of the crystal is an integer multiple of the period of the applied pulses;
for example in the 1:3 resonance peak (Fig.~\ref{kvfig2}(b)) three periods of the excitation correspond to one period of the traveling pulse in the crystal; so the excitation inflicts
a ``strong impulse" at every third period (whereas secondary, weaker impulses also occur in between, but they are not strong enough to initiate new traveling pulses in the crystal).
We conclude that the 1:$n$ resonance peaks correspond to traveling pulses in the crystal with periods equaling $n$ periods of the applied excitation.
This is confirmed in the plots of the
displacement waveforms of all beads during one full period (Fig.~\ref{kvfig2}(c) and Fig.~\ref{kvfig2}(d)).
The traveling pulse character of
these resonant responses of the crystal is also demonstrated
in Fig.~\ref{kvfig2}(e) and Fig.~\ref{kvfig2}(f),
where the effective phase variation (as defined above) of the individual bead responses is depicted as a function
of space (more precisely, of bead number). 

\begin{figure*}
\includegraphics[width=\textwidth]{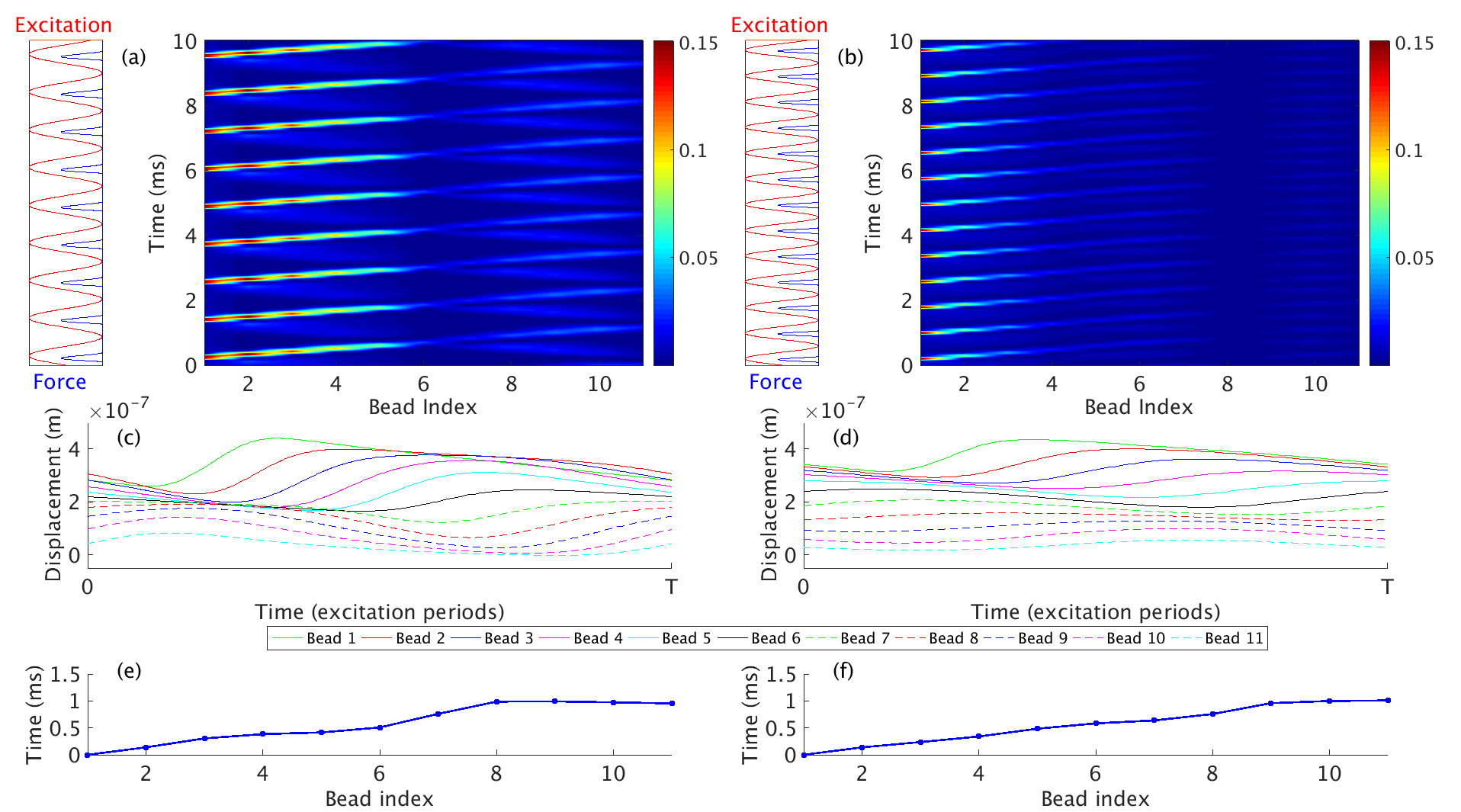}
\caption{(Color Online) (a),(b) Long-term, stationary spatio-temporal variation of the kinetic energy (scaled$\times {10^{7}}$) for (a) the second antiresonance (860 Hz) and
(b) the third antiresonance (1,260 Hz); the input harmonic displacement and applied force excitations for each case are also shown on the left of each plot.
(c),(d) Displacement waveforms of every bead during one full period of the crystal motion. 
(e),(f) Effective phase variation of the individual bead responses (computed by the time instants when each bead reaches maximum positive displacement
and when the leftmost bead does); this highlights the chimera-like, partially ``traveling pulse" and partially
``standing wave" nature of the long-term, stationary dynamics of the granular chain at antiresonances.}
\label{kvfig3}
\end{figure*}

Having explored the resonances of our nonlinear spectral picture, let
us now turn our attention to the corresponding {\em antiresonances}, corresponding to {\em local minima} of the (largest over one period) transmitted force to the right, immovable boundary.
As mentioned previously all antiresonances are associated with 1:1 periodic responses, and this is confirmed by the results depicted in Fig.~\ref{kvfig3}
where the second and third antiresonances are shown.
What is interesting is that these minima of the largest
transmitted force reside in the pass band of the crystal and alternate between resonances.
This strongly suggests that the observation of antiresonances is
not due to the incapacity of the granular crystal to transmit energy
(in the form of traveling solitary pulses) but rather it is
due to negative interference effects between left- and right-going traveling pulses;
in such cases one might expect to observe nonlinear
{\em standing waves}.
This is confirmed by the results of Fig.~\ref{kvfig3}, where negative interference effects between apparent traveling pulses 
can be observed, giving rise to peculiar periodic responses.
Indeed, it appears that there is a form of co-existence
of traveling pulses (as in the resonance cases discussed previously) and apparent nonlinear standing waves
in different parts of the crystal (traveling on the left side, standing
on the right side, the side of the immovable wall).
Such states of co-existence of different waveforms have been
dubbed chimeras in other contexts; see~\cite{chimera} for a relevant discussion.
This co-existence of apparently traveling and standing waves highlights the essentially nonlinear nature and the high complexity
of the long-term, stationary response, and is corroborated by the results presented in Fig.~\ref{kvfig3}(e) and Fig.~\ref{kvfig3}(f).
The negative interference of traveling pulses propagating in
opposite directions in the crystal would explain the minimization of transmitted force in antiresonances, 
and the ``low-intensity" impulses delivered by the exciting source in these cases. 
Such negative interference appears to lead
to the formation of standing wave breathing patterns, such as the ones
evident on the right side of the lattice in Fig.~\ref{kvfig3}(e) and Fig.~\ref{kvfig3}(f).

\begin{figure*}
\includegraphics[width=\textwidth]{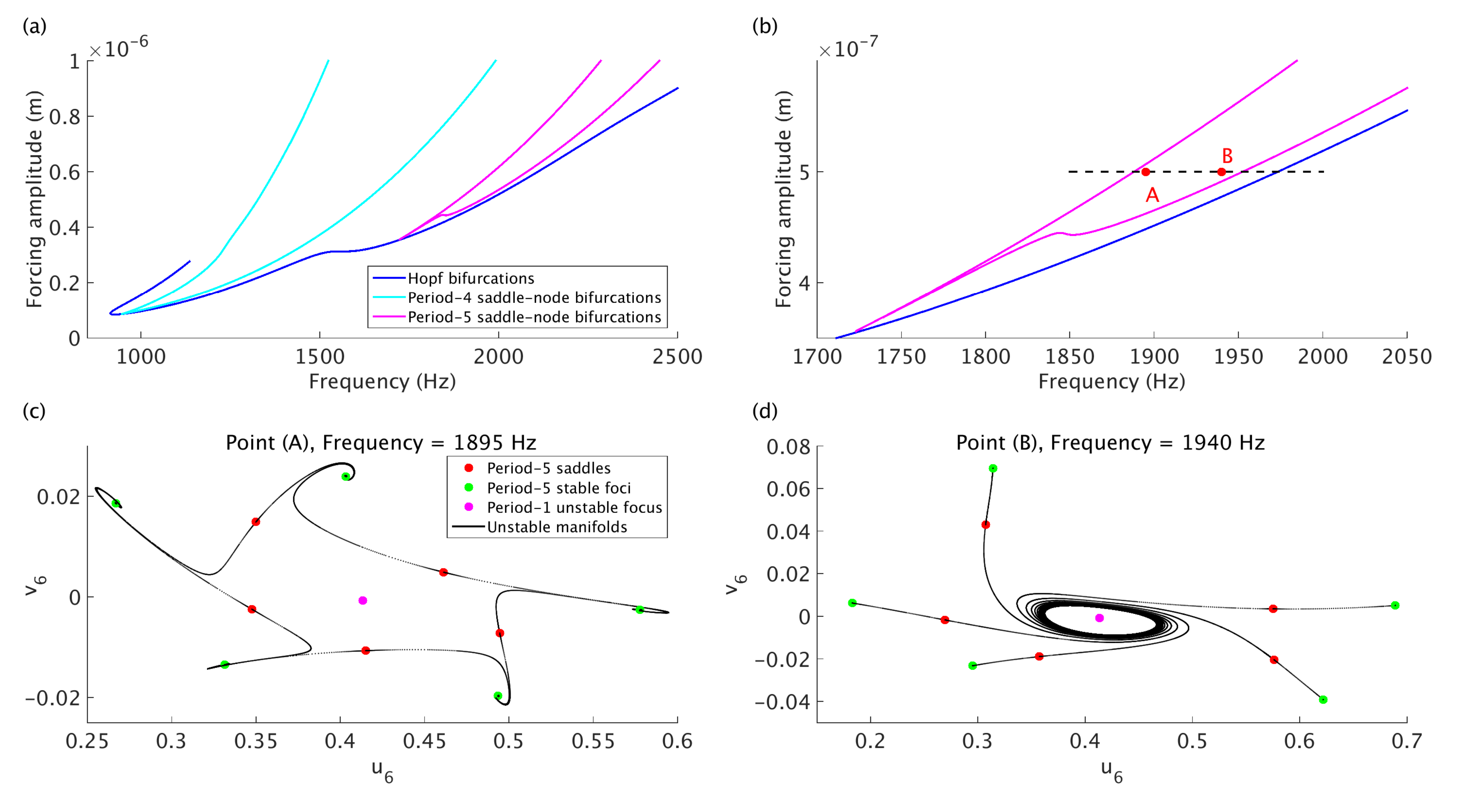}
\caption{(Color Online) (a) Partial two-parameter bifurcation diagram with respect
to the forcing amplitude and the frequency of the actuator. The curves denote
different bifurcations involving the nonlinear periodic solutions. More specifically:
(1) the continuation of a Neimark-Sacker
bifurcation of the period-1 solution (deep blue); and (2)
continuation of saddle-node bifurcations of the period-4 and period-5
solutions (cyan and purple, respectively).
The saddle-node bifurcations enclose
two resonance horns that are born on the Neimark-Sacker
curve when the Floquet multipliers cross the unit circle at the 4th and 5th roots
of unity respectively.
(b) Zoom-in of the period-5
resonance horn including a one-parameter cut across it.
(c) Poincar{\'e} cut showing that the unstable manifolds of the saddles at point A above, along with the 
stable period-5 solutions,  form a closed invariant set;
at the horn boundary this was an invariant circle on which the saddles and
the stable period-5s were ``born"  locked.
The axes here are the dimensionless position and velocity of the 6th bead. (d) Moving towards point
B in parameter space above, homoclinic crossings of the unstable and stable manifolds of the saddles have occurred
 and two attractors now coexist, a stable
invariant circle (a quasi-periodic solution)
and a stable node period-5 solution.}
\label{kvfig4}
\end{figure*}

All of the above analysis has been performed for a fixed value of
the forcing amplitude.
However, our system 
permits exploration of a wide parameter space by means of the
concurrent variation of both the forcing amplitude $A$ and
the forcing frequency $\omega$.
This more global perspective
of the system is provided in Fig.~\ref{kvfig4}.
Here, we use
the same color as before (in Fig.~\ref{kvfig1}) to denote
the bifurcation loci of the different branches.
For instance,
the isolas of the period-4 and period-5 solutions terminate
in pairs of saddle-node bifurcations (cf. Fig.~\ref{kvfig1})
that are followed, in a two-parameter bifurcation diagram as the amplitude $A$
varies, to produce resonance horns within which such solutions are contained in
the $(A,\omega)$ space.
We also highlight the locus of Neimark-Sacker period-1 bifurcations, giving rise to quasi-periodic solutions.
The tips of these resonance horns lie, in fact, {\em on the
Hopf (Neimark-Sacker) curve},
when the Floquet multipliers cross the unit circle at 4th and 5th roots
of unity respectively.
Needless to say that a host of other resonance horns (corresponding to every
rational number) and many secondary bifurcations, both local and global, that
cannot be feasibly computed, can also, in principle, be found in such a diagram
(e.g.~\cite{aronson1982,kevr1986,aronson1986}).
The bottom panels provide a sense of the variation of one of the system
Poincar{\'e} sections (here, by monitoring the plane of the displacement
and velocity of the 6th bead).
The existence of a stable
period-5 solution in both panels is illustrated by the green
points;  a coexisting unstable (saddle-type) period-5
 is shown by the red points, while the unstable
period-1 solution of the ``backbone" branch is shown by the isolated
point in the middle.
The saddle- and node- period-5 solutions were initially born in a saddle-node
on the torus; a remnant of this can be seen in that alternate sides of the
unstable manifold of the saddles are attracted to different stable period-5 points.
Global bifurcations, involving homoclinic manifold crossings and the associated
horseshoe dynamics (see e.g.~\cite{GuckHolmes}), also occur within these resonance horns;
while in the right panel a coexisting secondary
stable quasiperiodic attractor emerges in the form of an invariant circle.
In this last panel, one side of the saddle unstable manifolds is attracted by the stable period-5,
while the other side is attracted by this torus; this change clearly suggests that a global
bifurcation, namely 
a homoclinic interaction of the saddle invariant manifolds with each other, must occur 
along the path from point A to point B in parameter space.
This is but a glimpse of the considerable complexity of this highly nonlinear system and the
role of the nonlinear building blocks (stable and unstable periodic,
as well as quasi-periodic solutions) naturally arising in the nonlinear
analogue of the spectral picture we have outlined here.

{\it Conclusions and Future Challenges.} In summary, in the present work,
we focused on forced resonance spectra of highly nonlinear
homogeneous granular crystals in the complete absence of linear
acoustics (sonic vacua).
Despite the lack of sound propagation, in the sense
of classical acoustics, in the medium,
we observed that a purely nonlinear analogue of the spectrum arises.
This consists of harmonic and subharmonic responses to the external
boundary drive of the granular crystal, leading to a series of
peaks (resonances), reminiscent of the transmission frequency peaks
in a finite, linearly coupled lattice.
Here, the building blocks of the long-term, stationary responses are periodic
and quasi-periodic orbits, whose existence and stability we explored computationally.
Moreover, we illustrated how the periodic orbits corresponding to the
energy transmission (resonant) peaks emerge out of periodic oscillations within the chain 
that can be attributed
to highly nonlinear traveling pulses; the dips between them
(antiresonances) are
due to the emergence of a mix of apparent traveling pulses and
standing waves,
trapping the driver's energy
in the vicinity of the lattice edges (a surface
breathing mode, resembling a highly nonlinear analogue of
the well-known Tamm states~\cite{kivshar}).
These results were extended
into multiparametric variations (amplitude as well as frequency of the drive)
and their generic nature was confirmed.

This proof-of-principle demonstration paves the way for systematic future investigations
of such systems.
On the one hand, it would be interesting and relevant
to attempt to obtain a detailed experimental blueprint of the
full nonlinear spectrum; indeed, experiments guided by these predictions and 
confirming their validity
are underway, and will be reported elsewhere. 
On the other hand, numerous questions deserve further
exploration, regarding the specifics of corresponding spectral pictures
in systems with different types of nonlinear couplings, or  in
the presence of local
nonlinearities as, e.g., in the cradle model of~\cite{james},
as well as in the case of internal resonators~\cite{chiara}.
A systematic understanding of the modifications of the spectral
characteristics as $N$ progressively increases would be another
important aspect for future study.
Extending these considerations
to multi-dimensional systems with different geometric
characteristics~\cite{vak2,chiara2} would also represent
a significant generalization of the present work.
Such studies are presently in progress and will be reported
in future publications.

{\it Acknowledgments.}
AFV would like to acknowledge the support of MURI grant US ARO W911NF-09-1-0436. Dr. David Stepp is the grant monitor.
DP, MOW, PGK and IGK gratefully acknowledge the support
of US AFOSR
through grant FA9550-12-1-0332. PGK's work at Los Alamos is supported
in part by the Department of Energy.

\vspace{1\baselineskip}

{\it Appendix.} We consider a one-dimensional chain of spherical
elastic beads in Hertzian contact with no precompression:
\begin{widetext}
\begin{eqnarray}
m\frac{d^2u_1}{dt^2} &=& \frac{4}{3}E^*\sqrt{\frac{R}{2}}\left[(A'\sin{(2\pi ft)}-u_1)_+^{3/2} - (u_1-u_2)_+^{3/2}\right] \nonumber\\
\hspace{2em}&&+ D \left[(2\pi fA'\cos{(2\pi ft)}-\dot{u}_1) H(A'\sin{(2\pi ft)}-u_1) - (\dot{u}_1-\dot{u}_2) H(u_1-u_2)\right], \nonumber\\
m\frac{d^2u_i}{dt^2} &=& \frac{4}{3}E^*\sqrt{\frac{R}{2}}\left[(u_{i-1}-u_i)_+^{3/2} - (u_i-u_{i+1})_+^{3/2}\right] \nonumber\\
\hspace{2em}&&+ D \left[(\dot{u}_{i-1}-\dot{u}_i) H(u_{i-1}-u_i) - (\dot{u}_i-\dot{u}_{i+1}) H(u_i-u_{i+1})\right], \nonumber\\
m\frac{d^2u_N}{dt^2} &=& \frac{4}{3}E^*\sqrt{\frac{R}{2}}\left[(u_{N-1}-u_N)_+^{3/2} - (u_N)_+^{3/2}\right] \nonumber\\
\hspace{2em}&&+ D \left[(\dot{u}_{N-1}-\dot{u}_N) H(u_{N-1}-u_N) - (\dot{u}_N) H(u_N)\right], \nonumber
\label{eqn2}
\end{eqnarray}
\end{widetext}
where
\begin{eqnarray}
 m &-& \text{mass of each bead } (28.84 \text{ g}) \nonumber\\
 R &-& \text{radius of each bead } (9.525 \text{ mm}) \nonumber\\
 A' &-& \text{amplitude of the actuator } (5\times10^{-7} \text{ m}) \nonumber\\
 f &-& \text{frequency of the actuator } \nonumber\\
 D &-& \text{viscous damping coefficient} \nonumber\\
       &&\text{between beads } (100 \text{ Ns/m}) \nonumber\\
 E^* &=& \frac{E}{2(1-\nu^2)} \nonumber\\
 E &-& \text{Young's modulus } (193 \text{ GPa}) \nonumber\\
 \nu &-& \text{Poisson's ratio } (0.3). \nonumber
\end{eqnarray}
Define
\begin{eqnarray}
 A_0 &=& A' \text{ (original amplitude of the actuator)} \nonumber\\
 x_i &=& \frac{u_i}{A_0} \nonumber\\
 A &=& \frac{A'}{A_0} \nonumber\\
 \bar{t} &=& \tau t \nonumber\\
 \lambda &=& \frac{D}{m\tau} \nonumber\\
 \omega &=& 2\pi f, \nonumber
\end{eqnarray}
where
\begin{eqnarray}
 \tau &=& \sqrt{\frac{4E^*\sqrt{\frac{R}{2}A_0}}{3m}}. \nonumber
\end{eqnarray}
Using the variables defined above we end up with our nondimensionalized
system of equations
\begin{eqnarray}
\ddot{x}_i &=& (x_{i-1}-x_i)_+^{3/2} - (x_i-x_{i+1})_+^{3/2} \nonumber\\
\hspace{2em}&&+ \lambda \left[(\dot{x}_{i-1}-\dot{x}_i) H(x_{i-1}-x_i)\right. \nonumber\\
\hspace{2em}&&\hspace{1em}- \left.(\dot{x}_i-\dot{x}_{i+1}) H(x_i-x_{i+1})\right], \nonumber \hspace{50pt} (\ref{eqn1})
\end{eqnarray}
with $x_{N+1}=\dot{x}_{N+1}=0$ and $x_0=A\sin(\omega t)$.

\end{document}